\documentclass[conference]{IEEEtran}
\IEEEoverridecommandlockouts

\usepackage[frozencache,cachedir=.]{minted}
\usepackage{algpseudocode}
\usepackage[ruled,vlined,linesnumbered]{algorithm2e}
\SetAlCapNameFnt{\small}
\SetAlCapFnt{\small}
\SetAlgorithmName{Alg.}{Alg.}{List of Algorithms}

\usepackage{cite}
\usepackage{amsmath,amssymb,amsfonts,amsthm, latexsym}
\usepackage{graphicx}
 \graphicspath{{./img/}}
\usepackage{textcomp}
\usepackage{xcolor}
\def\BibTeX{{\rm B\kern-.05em{\sc i\kern-.025em b}\kern-.08em
    T\kern-.1667em\lower.7ex\hbox{E}\kern-.125emX}}
    
\usepackage{url}
\usepackage[english]{babel}
\usepackage[autostyle]{csquotes}
\MakeOuterQuote{"}

\usepackage{tipa}
\usepackage{soul}
\usepackage{enumitem}

\usepackage[font=scriptsize]{caption}
\usepackage{float}
\addto\captionsenglish{}
\usepackage{minted}

\IEEEoverridecommandlockouts
\IEEEpubid{\makebox[\columnwidth]{978-1-7281-2522-0/19/\$31.00~\copyright{}2019 IEEE \hfill}\hspace{\columnsep}\makebox[\columnwidth]{ }}

\begin{document}

\title{Blockchain-based Personal Data Management: \\ From Fiction to Solution}

\author{\IEEEauthorblockN{Nguyen B. Truong}
\IEEEauthorblockA{\textit{Data Science Institute} \\
Imperial College London, U.K \\
n.truong@imperial.ac.uk}
\and
\IEEEauthorblockN{Kai Sun}
\IEEEauthorblockA{\textit{Data Science Institute} \\
Imperial College London, U.K \\
k.sun@imperial.ac.uk}
\and
\IEEEauthorblockN{Yike Guo}
\IEEEauthorblockA{\textit{Data Science Institute} \\
Imperial College London, U.K \\
y.guo@imperial.ac.uk}
}

\maketitle

\begin{abstract}
The emerging blockchain technology has enabled various decentralised applications in a trustless environment without relying on a trusted intermediary. It is expected as a promising solution to tackle sophisticated challenges on personal data management, thanks to its advanced features such as immutability, decentralisation and transparency. Although certain approaches have been proposed to address technical difficulties in personal data management; most of them only provided preliminary methodological exploration. Alarmingly, when utilising Blockchain for developing a personal data management system, fictions have occurred in existing approaches and been promulgated in the literature. Such fictions are theoretically doable; however, by thoroughly breaking down consensus protocols and transaction validation processes, we clarify that such existing approaches are either impractical or highly inefficient due to the natural limitations of the blockchain and Smart Contracts technologies. This encourages us to propose a feasible solution in which such fictions are reduced by designing a novel system architecture with a blockchain-based "proof of permission" protocol. We demonstrate the feasibility and efficiency of the proposed models by implementing a clinical data sharing service built on top of a public blockchain platform. We believe that our research resolves existing ambiguity and take a step further on providing a practically feasible solution for decentralised personal data management.
\end{abstract}


\IEEEpeerreviewmaketitle

\section{Introduction} \label{INT}
In recent years, there have been a blooming volume of personal data from heterogeneous sources from social network to medical devices. The massive aggregation of personal data collected by centralised service providers (SPs) has raised critical privacy threats to billion clients (\textit{i.e.}, Data Owners (DOs)), particularly with the lack of mechanisms for effectively controlling and auditing personal data usage \cite{rs01}. The new General Data Protection Regulation (GDPR)\footnote{https://gdpr-info.eu} has taken a step further toward dealing with personal data abuses. As the GDPR legislation only provides data projection principles, solutions must be implemented as its complimentary, considering state-of-the-art technological resolutions. In this regard, the emerging Blockchain (BC) and Smart Contract (SC) technologies, which play as the backbone of a variety of decentralised applications and services, are expected to be a prospective solution. Thanks to the features such as decentralisation, immutability, trace-ability, and transparency; a BC-based personal data management system could \textit{(i)} bring full control back to data owners; \textit{(ii)} provide secure and efficient access control mechanisms for data usage; \textit{(iii)} cater auditability and provenance tracking of data access; and \textit{(iv)} decentralise the system by removing trusted intermediaries and prevent the single-point-of-compromise and -failure.

A number of research articles have stated technical potentials of blockchains in managing personal data \cite{ref32, ref24, ref25, ref33, ref34}; however, most of them only provided preliminary methodological exploration or conceptual models without analysing and implementing a BC-based personal data management system in details. Such systems adopt an undeniable holistic architecture of decoupling the BC, which is for accounting and auditing data access, from a storage layer, which physically stores data. Unfortunately, there are some flaws in existing BC-based personal data management systems which have been disseminated in the literature when actualising the architecture. Such fictions are consequences of a system design that let a BC network fully handle data access requests which directly interacts with an off-chain storage. This design includes mechanisms which are theoretically achievable. However, by thoroughly inspecting consensus protocols and transaction validation processes, we figure out that those mechanisms are either technologically unfeasible or critically inefficient due to nature limitations of the BC and SC technologies.

As a motivation, we envisage a decentralised BC-based personal data management system in which such fictions are eliminate by re-designing the existing system architecture. In the proposed system, a BC is only dedicated to authenticate and authorise permission of requests; data storage and retrieval are performed directly between a requester and a provider. For this purpose, we introduce a novel authentication and authorisation mechanism for personal data query leveraging the concept of decentralised \textit{"access token"}, inspired by the centralised delegated authorisation approach introduced in OAuth\footnote{https://oauth.net/2} standardisation. The feasibility and efficiency of the proposed models are then demonstrated by an Ethereum-based clinical trials management system.

The rest of the paper is organised as follows. Section II briefly presents relevant knowledge and related work. Section III clarifies fictions with analysis on the existing approaches. Section IV describes our proposed solution. Section V is dedicated to the solution demonstration. Section VI concludes our work with future research directions.
\section{Background and Related Work} \label{3RelatedWork}

\subsection{Blockchain Technology}
BC technology comprises of diversified techniques including distributed computing, computer network, database, cryptography, privacy and security. The goal of BC technology is to create and maintain a distributed appended-only database constituted from a chain of blocks (so-called BC) in which each block contains a list of transactions organised in Merkle tree, and is linked with the previous block by cryptographic block hash. A BC plays a role of a distributed ledger maintained by peers in a trustless peer-to-peer network in a decentralised manner for a variety of business logic. Literally, a BC is resistant to data modification as altering information in a block requires to negate all hashes in the previous blocks which is replicated in all nodes in the network, breaking the consensus among them. This implies an attacker has to take control over $50\%$ nodes in the network, which is supposed to be impossible \cite{rs05}. BC's features including tamper-resistance, transparency and traceability make it applicable to auditing and accounting digital assets, particularly cryptocurrencies \cite{ref05}.

A consensus protocol, as a core of BC technology, ensures all peers in a trustless environment agree on which transactions are legitimate to be added to a BC; thus, synchronises the network to maintain a consistent and unique BC among peers \cite{ref06, ref07}. Proof of Work (PoW) is the most popular consensus protocol currently used in Bitcoin and Ethereum in which dedicated nodes (\textit{i.e.,} miners) race to produce \textit{hashcash} as an entitlement to add a new block to the chain for some incentives \cite{ref08}. The computation-intensive for solving the hashcash puzzle appears as a critical bottleneck of PoW; therefore, alternatives have been proposed to mitigate this deficiency including Proof of Stake (PoS) \cite{ref09}, Proof of Authority, and Byzantine fault-tolerant (BFT) variants \cite{ref11}. Still, these consensus protocols impose severe drawbacks resulting in limited usage compared to the PoW \cite{ref07}.

\subsection{Smart Contracts}
A SC is a self-executing program implementing business logic of a decentralised service using BC. As being deployed into a BC network, it is a decentralised automation that verifies and facilitates transactions for enforcing terms and conditions in the business logic written in form of computer code in the SC (as clauses in a contract, hence the name) \cite{ref13}. Once a party submits transaction to invoke a SC, contract clauses written in the SC is automatically enforced without the need for a central enforcement authority. Any BC framework provides facility for SCs from a simple stack-based scripting system (\textit{e.g.,} in Bitcoin) to a Turing-complete programming language (\textit{e.g.,} Solidity in Ethereum). Ethereum is the most well-known platform offering programmable capability with Ethereum Virtual Machine (EVM) on which code of arbitrary algorithmic complexity is compiled and executed \cite{ref15}.

\subsection{Personal Data Management: Scenarios and Challenges}
\begin{figure}[!htbp]
\centering
	\includegraphics[width=0.4\textwidth]{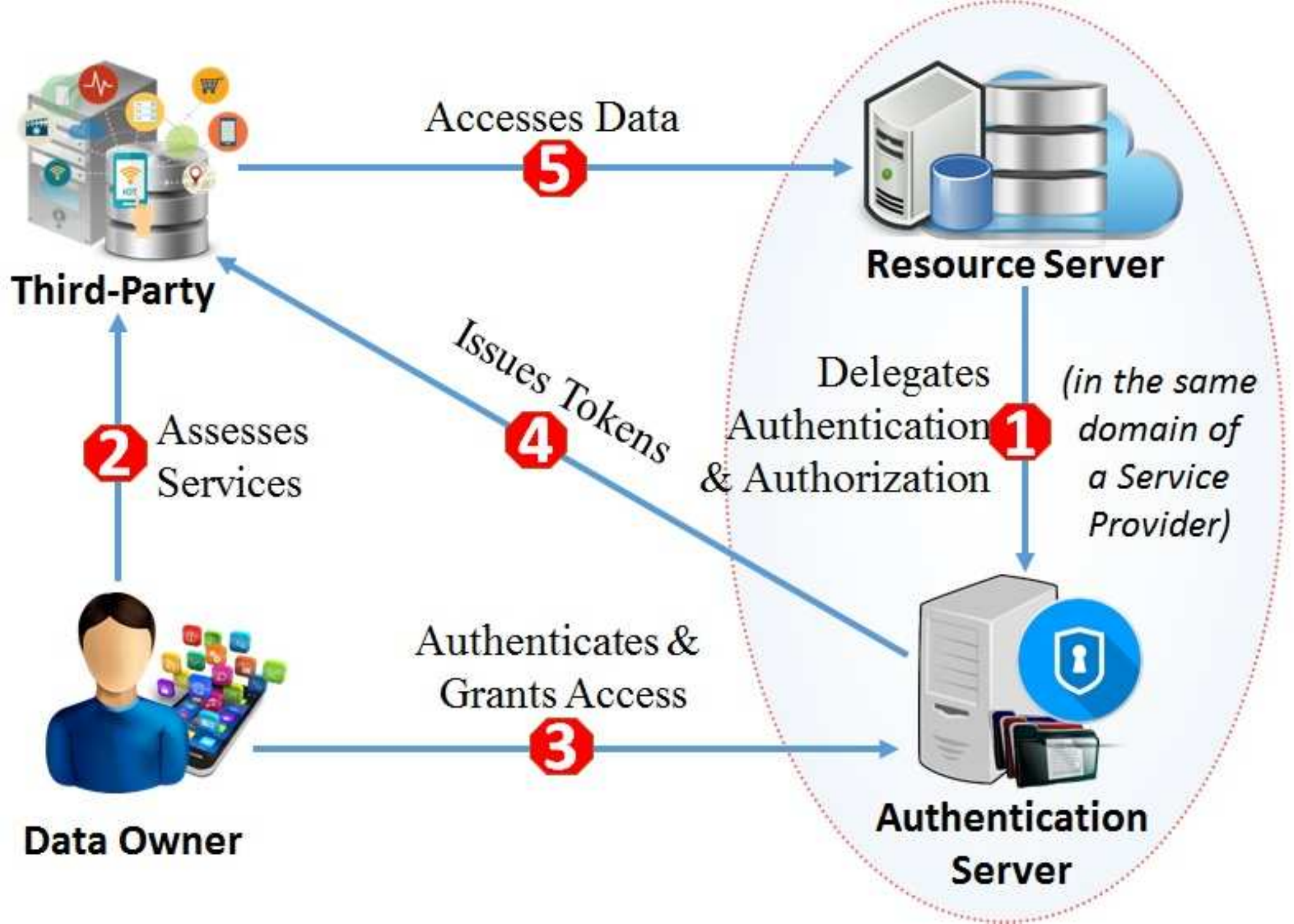}
	\caption{Personal data management and sharing in conventional client-server architecture leveraging a delegated authentication server}
	\label{fig1}
\end{figure}

Generally, a DO grants a set of permissions for collecting and processing data once using a service provided by an SP. When the DO starts using a service provided by a Third-party (TP), the TP asks for permission to access DO's data that is collected and managed by the SP. A majority of traditional personal data management and sharing mechanisms are implemented under a centralised client-server architecture leveraging a delegated authentication and authorisation server (Fig. \ref{fig1}), following the $OAuth2$\footnote{https://oauth.net/2/} access delegation standard. This approach allows DOs to share their personal data managed by an SP in a simplified fashion with a single log-on.

As illustrated in Fig. \ref{fig1}, to grant permission for the TP, the DO logs on the SP's service and accepts access request from the TP (step 3). The Authentication server authorises the permission by providing \textit{access tokens} (step 4). The TP then accesses desired data from a Resource Server (RS) with the tokens as proof of permission (step 5). However, the centralisation of this approach demonstrates limited transparency and a lack of trust, which consequently poses variety of critical concerns \cite{ref40}. This is because personal data is under full control of an SP - the only authority to \textit{(i)} authenticate and authorise participants; and \textit{(ii)} manage data access and provenance. In this respect, personal data management would be a well-suited scenario for BC technology to shine \cite{ref32}.

\subsection{Related Work}
Significant efforts have been carried out that leverage BC for storing crucial information with decentralised automation (\textit{i.e.}, SCs) for asset auditing, access and permission control while eliminating the reliance on an intermediary. For instance, BC is used for provenance tracking in logistic and supply chain in \cite{rs06} in which business operations are integrated and logged in immutable distributed ledgers so that product data can be tracked and traced in a decentralised manner. In \cite{ref21} cloud data provenance comprising of full history of operations upon a data object in a cloud server is recorded in a BC; thus, preventing from unauthorised access/modification to the provenance data. BC is also used for brokering services in which a SC is deployed playing as an automated negotiator between clients and an SP \cite{rs07}. Blockstack project \cite{ref22, ref23} replaced traditional centralised domain name system (DSN) with a decentralised one. Its novel system architecture decoupled a BC (which stores only domain registration and hash of off-chain data) from an off-chain cloud data storage (which stores data payload and owners' digital signatures). This separation of on-chain operational management and off-chain data storage shed the light on other studies that utilise BC for general-purpose distributed data storage. For example, two data management platforms for the Internet of Things (IoT) have been proposed in \cite{ref24, ref25}. Data generated from IoT devices is stored in a Distributed Hash Table (DHT) whereas pointers to the data (\textit{i.e.,} data hash) are recorded on-chain. DHT nodes physically store data and listen to transactions confirmed by BC to store/send data from/to IoT devices accordingly. BigchainDB \cite{ref26} targets a general-purpose distributed storage system that replaces the P2P communication in a traditional distributed database by Tendermint \cite{rs09}, a BFT consensus protocol for networking and consensus. Nodes in BigchainDB store data in their local MongoDB databases while operations upon the local databases (\textit{e.g.,} replicas, replication factors) are controlled by a federation of nodes in a permissioned BC fashion using Tendermint.

Regarding personal data management, authors in \cite{ref32} proposed a privacy-preserving platform for data storage and sharing in which actual data is stored off-chain while data usage agreement, access control policy and data pointers are on-chain. Public-key cryptography is also exploited defining the role of a DO and an SP in a dataset for identity verification and authentication in a data query request. The BC platform then authorises data query requests by looking for necessary information (\textit{e.g.,} access control policy and permission grant) published in BC. This work has been adopted for numerous scenarios including IoT data \cite{ref24, ref25} and medical records \cite{ref33, ref34, ref35}. Similarly, a personal files management platform deployed on top of the Ethereum framework is proposed in which attributed-based encryption (ABE) policy is recorded on-chain for providing fine-grained access control \cite{ref30}. Actual files are stored using Interplanetary File System (IPFS), a distributed file storage system \cite{ref31}. In these works, a BC plays as an intermediary between DOs and SPs and an off-chain data storage carrying out missions of authenticating, authorising access queries as well as storing and delivering data. Further analysis of this approach will be provided in the next section.
\section{Fiction Analysis on Existing BC-based Personal Data Management Solutions}
\subsection{Current Solution Approaches}
\begin{figure}[!htbp]
\centering
    \includegraphics[width=0.45\textwidth]{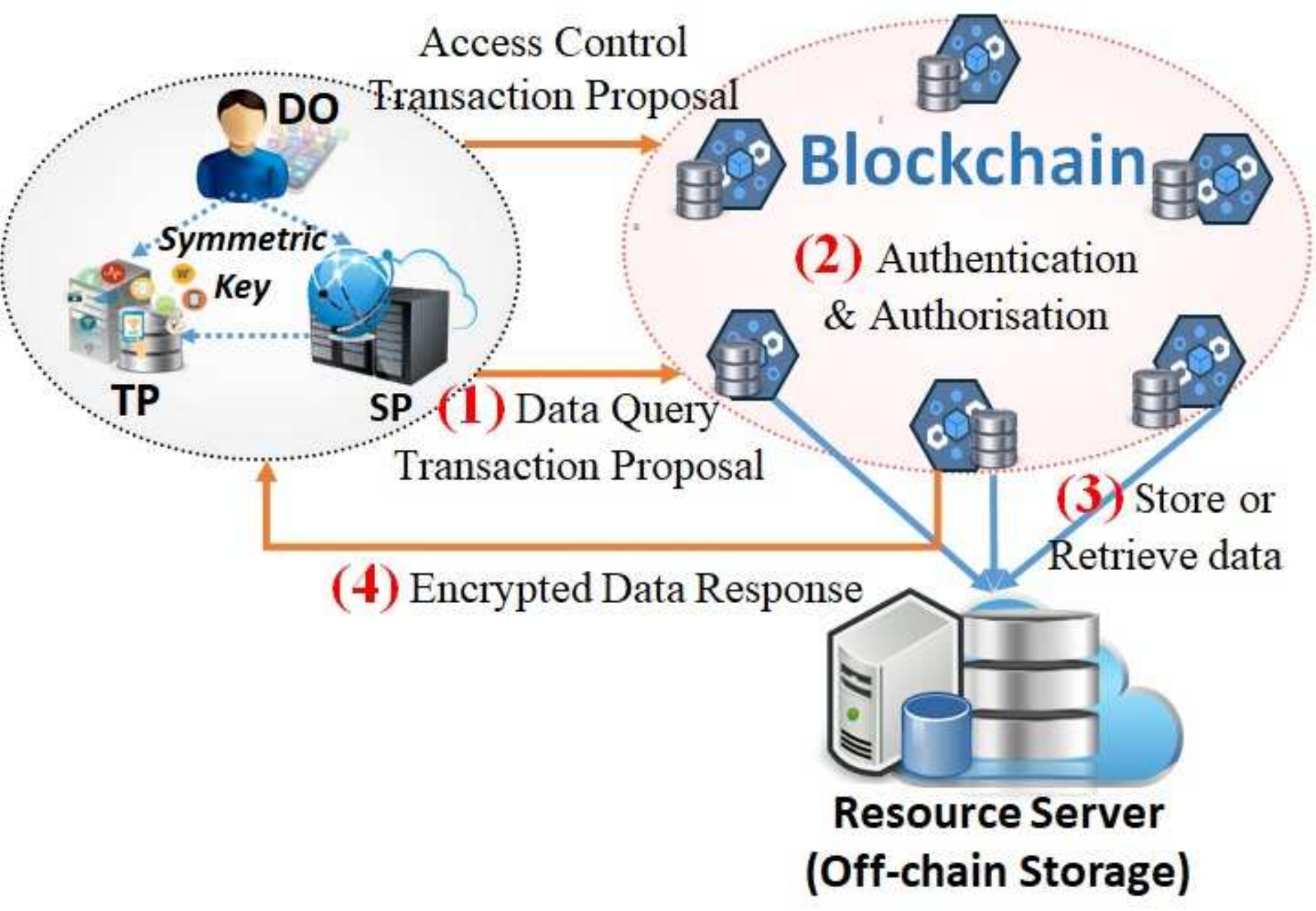}
	\caption{System architecture of the existing BC-based personal data management}
	\label{fig2}
\end{figure}

Existing approaches on BC-based personal data management adopted a system architecture constructed from a BC and an off-chain data storage, as shown in Fig. \ref{fig2}. Accounting information such as permission grant and data usage policy, which requires to be immutable and traceable, was recorded on-chain. Actual data, instead, was stored off-chain. The separation is a must as it is impractical to store personal data on a BC; instead off-chain storage offers significantly higher efficiency and better scalability \cite{ref22, ref23, ref26}. Only references to the actual data (\textit{e.g.}, hash pointers) were recorded in a ledger.

The general idea of existing solutions is that a BC acts as an automated access control manager ensuring DOs have full capability to control their personal data. Hence, the BC handles two types of requests: Data Query Transactions (DQ-Tran) for data upload and retrieval, and Access Control Transaction (AC-Tran) for managing data usage policy. Handling of AC-Tran is straightforward: a party proposes a transaction comprising of an ID, a digital signature and a data usage policy for a dataset (\textit{e.g.}, grant a consent, revoke a consent, and update permissions) to the BC. The BC verifies the signature to see \textit{(i)} whether the requester has a right to impose the usage policy (\textit{e.g.}, the requester is the owner or service provider of the dataset); and \textit{(ii)} if it is the case, the BC accomplishes the AC-Tran by appending the usage policy to a corresponding ledger. Regarding a DQ-Tran, the current solutions implement a generic four-step procedure as illustrated in Fig. \ref{fig2}:
\begin{enumerate}[label=(\roman*)]
    \item \textbf{Transaction Proposal}: A party proposes a DQ-Tran to a BC comprises of a digital signature and an operation parameter (\textit{e.g.,} \textit{READ} or \textit{WRITE}). If the operation is \textit{WRITE}, a data payload is then provided along.
    \item \textbf{Authentication \& Authorisation}: The BC verifies the signature and authorises whether the party has permission to conduct the requested operation. The authorisation can be done by inspecting a corresponding access control policy recorded in a distributed ledger.
    \item \textbf{Data Store/Retrieval}: Once the party get authorised, the BC looks up the ledger for the $data\ pointer$ of the requested dataset. The BC then uses this $data\ pointer$ to upload or retrieve the dataset which is physically stored in a off-chain Resource Server (RS).
    \item \textbf{Data Query Response}: The BC finally returns the data obtained from step (3) to the requester if the requested operation is \textit{READ}. If the operation is \textit{WRITE}, the BC may return an ACK message, depending on each system.
\end{enumerate}

\subsection{Fiction Analysis}
The generic procedure for DQ-Tran imposes some fictions clarified in details as follows:
\subsubsection{Overhead Fiction}
The overhead fiction comes from a system design that let a BC network fully handle data query requests without fully examining the underlying operations of the BC framework, particularly the consensus protocol, the transaction validation and ledger update processes. In general, a transaction life-cycle follows the 4-step procedure:
\begin{enumerate}[label=(\roman*)]
    \item \textbf{Transaction Proposal}: The client app proposes a transaction to invoke a SC for querying personal data, along with the party's digital signature for authentication.
    \item \textbf{Broadcast and Queuing}: The transaction proposal is broadcast to the BC network and queued at the transaction pool (\textit{i.e.,} $MEMPOOL$) in each miner.
    \item \textbf{Block Mining}: Miners selects the proposal to include in a block (with other transaction proposals), it is validated and executed. \textit{Output} of the execution with the \textit{transaction proposal} are combined together forming a \textit{transaction} which is then put into a pending block. The miners start mining the block. A miner who wins consensus race (\textit{e.g.,} solves a \textit{hashcash} in PoW) seals the block using \textit{block hash} and broadcast it to the network.
    \item \textbf{Block Validation and Confirmation}: Eventually all full-nodes in the BC network receive the new block after a certain bound $\Delta$. Upon receiving the block, all transactions in the new block are validated. If each transaction is valid, each full-node \textit{locally executes} the transaction with its local BC state $S$ (line 5 in Alg. \ref{alg1}) and \textit{compares} the local result (\textit{i.e.,} a new BC state $S'$) with the result sealed in the new block (line 6). If the results are the same for all transactions; the node synchronises the local BC with block $B$.
\end{enumerate}

\begin{algorithm}
    \footnotesize
    \SetKwInOut{Input}{Input}
	\SetKwInOut{Output}{Output}
	\Input{new block $B$; BC state $S$}
    \Output{$true/false$}
	\BlankLine
	\For {\textbf{each} (transaction $T$ in block $B$)} {
	    \If {(\textbf{Validate}($S$, $T$) == false)}
	    {
	        \textbf{halt}; \\
	        \textbf{Return} false;
	    }
	    $S' \leftarrow$ \textbf{\textit{Execute}}($S$, $T$) \Comment{Obtain new BC state} \\
	    \If {($Merkle\_root(S')$ $\neq$ $Merkle\_root(B(T))$)}
	    {
	        \textbf{halt}; \\
	        \textbf{Return} false;
	    }
	    \textbf{\textit{Replace}} $S \leftarrow S'$
	}
	\textbf{Return} true;
	\caption{$Block Validation$ validates all transactions in the block and only updates the local BC if the results are the same.}
	\label{alg1}
\end{algorithm}

Let take a DQ-Tran created by a party whose permission is already granted for instance. As depicted in the life-cycle, the DQ-Tran is executed by some miners in step (3) and by all full-nodes in the BC network in step (4). Consider a weakly synchronous BC network consists of $N$ full-nodes with propagation delay is bounded on $\Delta$ seconds. Assume that in a traditional centralised approach, a RS handles $k$ requests per second (RPS) on average. As each full-node locally queries off-chain data for transaction validation; as a consequence, the RS in a BC-based system has to handle \textit{at least}:
\begin{equation}
	\label{eq_1}
    k \times \frac{N}{\Delta} \text{ RPS on average}
\end{equation}

To illustrate the consequence of the overhead, let us consider the Ethereum \textit{main-net} consisting of more than $25000$ nodes with the propagation delay is less than $10$ seconds\footnote{https://ethstats.net}. Therefore, if a personal data management system deploys on top of the Ethereum platform, a RS has to approximately handle more than $2500$ times of the current throughput on average. If the system is deployed on top of the Bitcoin network, there are around $9500$ reachable nodes\footnote{https://bitnodes.earn.com} with the propagation delay is measured at around $9$ seconds \cite{ref08}; thus a RS has to handle about $1050$ times higher than the current capacity. This is obviously unacceptable.

Moreover, this system design imposes threats on Distributed Denial of Service (DDoS) attack as it extremely amplifies the number of requests to the RS in a short duration of time. Furthermore, a number of the requests have to queue at the RS' buffer; thus there is high probability that some requests are dropped due to limited-size queue. This results in lowering down system efficiency, even not reaching consensus, as well as increasing the risk of buffer overflow attacks.

\subsubsection{The Oracle Problem Fiction}
The transaction validation expects any full-node produces same results when locally executing a SC; otherwise no consensus can be achieved. This implies the SC is \textit{deterministically} executed. Generally, SCs acquire information from ledgers - consensuses maintained by all nodes in the network. There might be a problem if a SC requires information from outside world, which might impose room for ambiguity. This also turns the whole BC into a centralised system as it relies on a reliable source of information. For instance, Ethereum does not provide API calls capability or built-in probabilistic functions except \textit{block hash} and \textit{timestamp}. Hyperledger Fabric offers more flexibility as its chaincodes can be implemented in multiple high-level programming languages with no restriction on determinism. However, making API calls to outside world always comes with high caution to ensure designated peer nodes retrieve exactly same responses. The need for acquiring external information to execute a SC poses a severe technical barrier to the BC technology, which is known under the term: \textit{"Oracle problem"} \cite{rs03}.

The current approaches on personal data management obviously impose the Oracle problem. This is because BC nodes are required to access off-chain data for checking data integrity in order to prevent a malicious miner from returning false data to a requester. For instance, hash of the requested data is recorded on-chain. Once receiving a new block from a miner, each full-node executes a corresponding SC to query the data from a RS then compares the hash of the obtained data with the hash provided in the new block for validating data integrity. Indeed, there are solutions for the Oracle problem by introducing decentralised trusted providers who feed required data into SCs. These data providers are called Oracles. For example, Oraclize\footnote{https://docs.oraclize.it/} deals with the issue by \textit{(i)} obtaining data from an external source and delivering the data to the corresponding SCs \textit{(ii)} providing an \textit{authenticity proof} ensuring data integrity. Currently, there is no solution tackling the Oracle problem on personal data management. Moreover, with the existing solution approaches, such Oracles have to handle bursts of request overheads from all nodes in a BC network, resulting in dramatically deteriorating system performance.

\subsubsection{Data Encryption and Fine-grained Access Control Fictions}
As not all nodes in a BC are trustworthy, personal data must be encrypted before carrying out a DQ-Tran. This is because a DQ-Tran requires any full-node to either obtain data payload (in $WRITE$ operation) or retrieve data (in $READ$ operation). Therefore, a shared encryption key must be established among parties in advance. Certainly, if an SP requires a distributed off-chain storage platform (\textit{e.g.,} IPFS \cite{ref31}), data encryption is a must as storage nodes might be malicious and insecure. However, if an SP uses a secure trusted DBMS (\textit{e.g.,} Oracle and MongoDB) or a cloud storage service (\textit{e.g.,} S3, AWS and Azure), the system design appears to be expensive and inadequate. The current approaches inflict consequences on system performance, functionality and access control due to the following reasons:
\begin{enumerate}[label=(\roman*)]
	\item \textbf{Expensive encryption scheme}: Data encryption and decryption always come at high cost; particularly with large volumes of data or with data streams.
	\item \textbf{Content-query not supported}: As data is in encrypted form, data query is limited to only obtain data at a whole; content-queries are not supported. Thus, these solutions are applicable to data storage only.
	\item \textbf{Coarse-grained Access Control}: The encrypted form of personal data severely limits sharing capability at a fine-grained level and imposes a painful revocation scheme \cite{rs04}. Homomorphic Encryption can be used for computation (\textit{e.g.,} query data) on cipher-text but it is extremely expensive \cite{rs08}. Flexible encryption schemes like attribute-based encryption (ABE) \cite{rs04} might also be a remedy; however, they rely on a trusted key generator, which require to be customised for a BC-based system.
\end{enumerate}
\section{Proposed Solution}
In this section, we propose a novel solution to eliminate the aforementioned fictions. We consider scenarios that an off-chain storage RS follows the "honest-but-curious" security model whereas SPs and TPs might be malicious. This means the RS honestly performs all required protocols although it might be curious about the results. A legacy DBMS or a cloud storage service satisfies this assumption.

\subsection{System Design and Procedures}
\begin{figure}[!htbp]
\centering
	\includegraphics[width=0.45\textwidth]{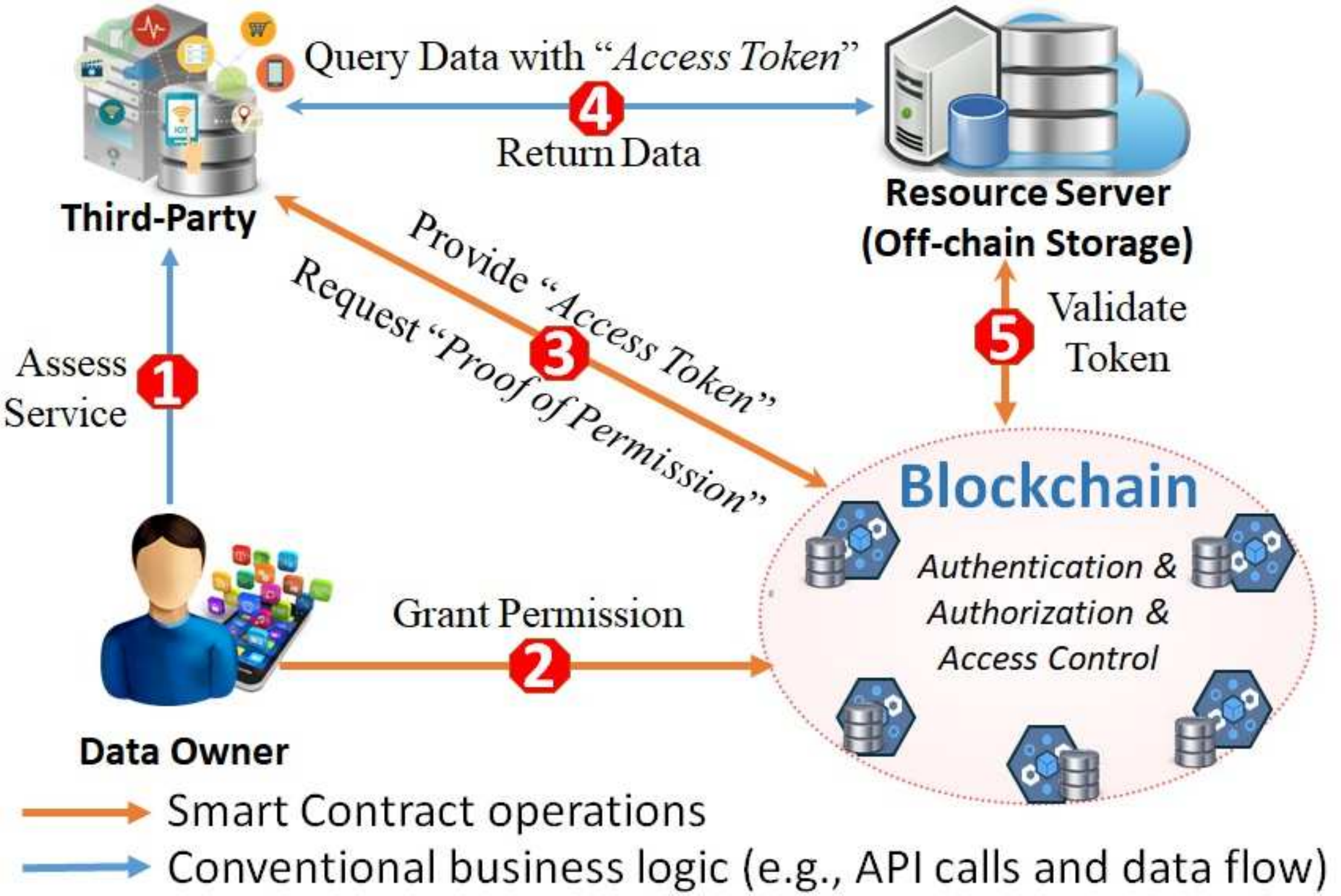}
	\caption{System architecture and procedures of the proposed solution}
	\label{fig3}
\end{figure}

The key idea of the proposed solution is to let a BC play only two roles: \textit{(i)} a delegated authentication \& authorisation server; and \textit{(ii)} an immutable logging system. Other personal data management business logic such as Data Storage and Retrieval remain the same as in traditional client-server approaches (Fig. \ref{fig3}). For this purpose, the BC issues an \textit{access token} as "proof of permission" showing that a party has been granted to access the designated data. The party then provides the \textit{access token} as a parameter in API calls to access data stored in a RS, similar with conventional client-server approaches. As a result, the \textit{Overhead} and the \textit{Oracle Problem} fictions are eliminated. Furthermore, this system design does not let any nodes in the BC network access personal data, along with the assumption of an honest and secure RS, the \textit{Data Encryption} fiction is also removed.

As illustrated in Fig. \ref{fig3}, the overall system procedure is as follows: Once a DO uses a service provided by a TP (step-1) and the TP requires to access DO's data stored in a RS, the DO grants consent to the TP by carrying out a AC-Tran to a BC (step-2). The BC authenticates and executes the AC-Tran, updating the access control policy recorded in a distributed ledger to reflect the new consent. Until this step, the proposed solution is completely the same with the existing approaches. The difference is from step-3 onward when a party requests to query data by performing a DQ-Tran. In our solution, the BC neither directly retrieves off-chain data nor returns data back to the requester; instead it provides an \textit{access token} as "proof of permission" (step-3) and let the party directly query data from a RS using the provided \textit{access token} (step-4). The RS validates the data query request (including the \textit{access token}) with the BC (step-5) before returning the requested data in case the request is valid. The honest RS assumption is utmost important to ensure that the RS genuinely executes the validation on whether an \textit{access token} is valid and used by a corresponding authorised party. The system design requires no changes in the SP's and RS's business logic. The only requirement is the involvement as clients of the BC system to perform required SCs' functions.

\subsection{Distributed Ledgers Data Structure}
Distributed ledgers recorded onto BC contain necessary information for personal data management business logic. This information must be specified as it is the main content of the ledgers and the core of any BC-based applications. For easy illustration, Ethereum is used as the reference BC framework, thus, the distributed ledgers comprise of time-stamp sequenced records of all state transitions. A state is in form of are in form of \textit{key-value} pairs, and a state transition records all changes to the key-value pairs as a result of a transaction. For the proposed personal data management, two ledgers are constructed namely $access\_ledger$ and $token\_ledger$ containing information for access control and \textit{access token} management. Data models for $access\_ledger$ and $token\_ledger$ are described in Listing \ref{lst1} and Listing \ref{lst2}, respectively.

\begin{listing}
\begin{minted}[frame=single,
               framesep=1mm,
               linenos=true,
               xleftmargin=15pt,
               fontsize=\footnotesize,
               tabsize=10]{json}
"access_ledger": {
    "key": {
      "DO_ID": "address",
      "SP_ID": "address",
      "TP_ID": "address"
    },
    "value" {
      "data_pointer": "string",
      "data_hash": "bytes32",
      "access_token": "bytes32",
      "permission": "uint8"
}}
\end{minted}
\centering
\caption{$access\_ledger$ data model in JSON format} 
\label{lst1}
\end{listing}

\begin{listing}
\begin{minted}[frame=single,
               framesep=1mm,
               linenos=true,
               xleftmargin=15pt,
               fontsize=\footnotesize,
               tabsize=10]{json}
"token_ledger": {
    "key": {
      "access_token": "bytes32"
    },
    "value" {
      "DO_ID": "address",
      "SP_ID": "address",
      "TP_ID": "address",
      "issued_at": "uint256",
      "status": "bool",
      "permissions": "uint8",
      "expires_in": "uint", 
      "refresh_count": "uint"
}}
\end{minted}
\centering
\caption{$token\_ledger$ data model in JSON format.} 
\label{lst2}
\end{listing}

The $access\_ledger$ is for authorisation and access control mechanisms contains information about DO, SP and TP identities (as "key" in key-value pair), and $data\_pointer$, $data\_hash$,  and  $permission$ as a simple data usage policy specifying CRUD operations (as "value" in key-value pair). The $token\_ledger$ is dedicated for managing $access\_token$ with information about DO, SP and TP identities and associated meta-data. The $access\_ledger$ is connected with the $token\_ledger$ by referring the \textit{access token} in its value for additional access control and audit tasks.

\subsection{Smart Contract Functionality}
\begin{figure}[!htbp]
\centering
	\includegraphics[width=0.48\textwidth]{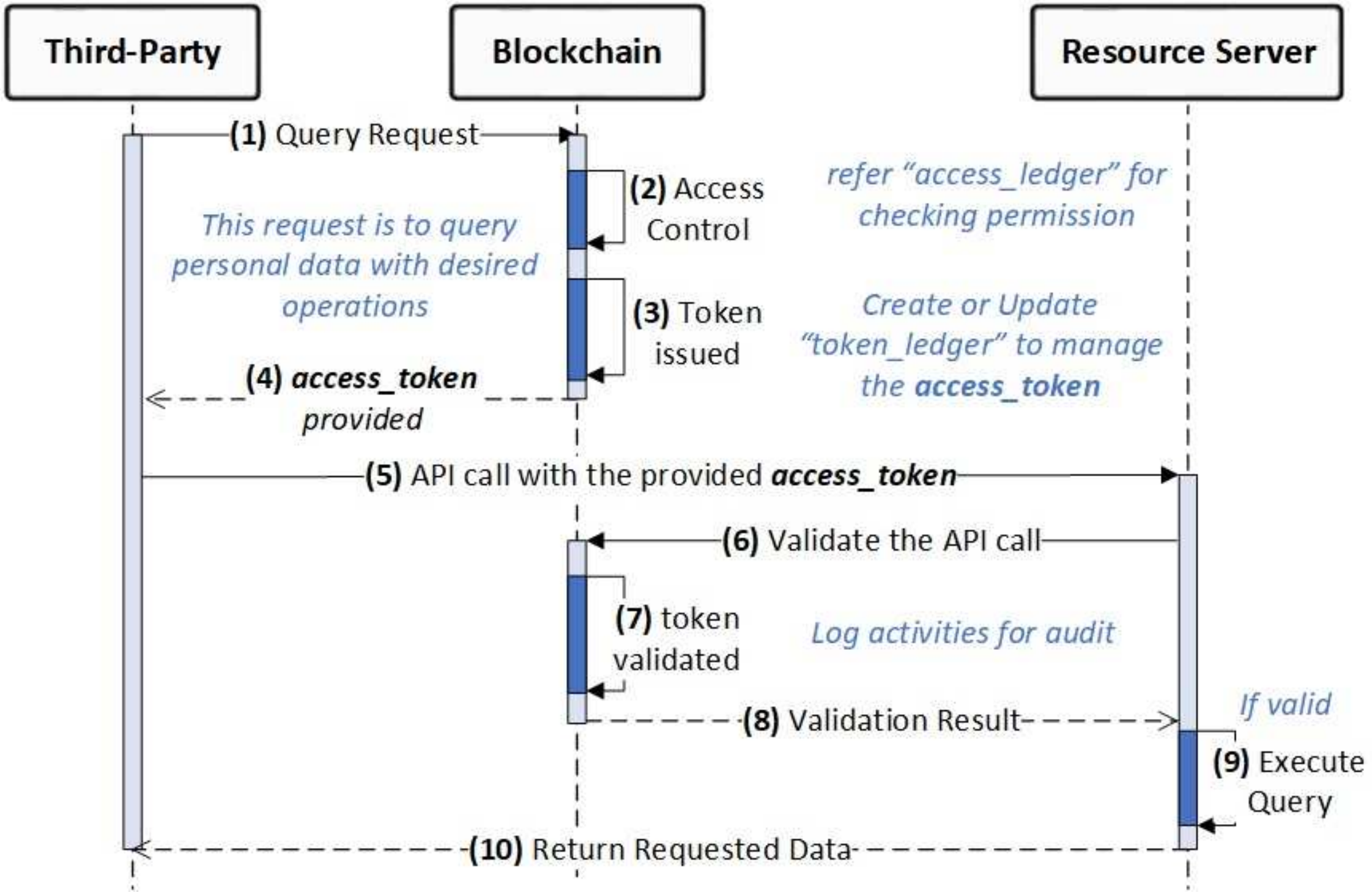}
	\caption{Data Query sequence diagram}
	\label{fig4}
\end{figure}

Variety of functions should be implemented in a SC for accounting access control and data query activities. Many functions in the proposed system are similar with ones in existing approaches, particularly the functions for performing AC-Trans such as \textit{Registration}, \textit{Permission Grant}, \textit{Permission Revocation}, and \textit{Permission Check}. Additional functions are implemented as a result of the introduction of the "proof of permission" recorded by the $token\_ledger$. As a reward, there are no more functions implementing DQ-Trans tasks (\textit{e.g.,} upload and retrieve data to/from an off-chain storage system). Fig. \ref{fig4} depicts the data query sequence diagram in which step 3-10 demonstrates the differences of the propose solution compared to the existing solutions. As illustrated in Fig. \ref{fig4}, if a query request is accepted (step 2), a new record of $token\_ledger$ as "proof of permission" is generated (or updated in case it is already existed) (step 3) before returning to the TP (step 4). The creation of the record is straightforward as it generates deterministic meta-data except the \textit{access token} value (\textit{i.e.,} the $key$ in the key-value pair "$token\_ledger$"). The \textit{access token} should be (pseudo)random and unique, and a trick is used to generate an unique pseudo-random token in Ethereum as follows:
\begin{equation}
	\label{eq_2}
    access\_token = hash(Parties\_IDs + Block.Hash)
\end{equation}

Alg. \ref{alg2} depicts the $Validation$ function invoked when a party call an API to the RS providing an \textit{access token}, an operation $op$, party's identity (\textit{i.e.,} $address$) $pk$ and a digital signature $t$ as parameters (step 5-8 in Fig. \ref{fig4}). The algorithm firstly validates the identity of the API requester using an signature verifying function $\mathcal{V}$ (line 2). If the request is from DO or SP then there is no further check for the \textit{access token}; only meta-data for the token in $token\_ledger$ is updated (line 5-8). Otherwise, the validation is conducted by inspecting the meta-data (line 9-12) before updating the $token\_ledger$ (line 12, 13). This function ensures that only API calls with valid \textit{access token} are executed (step 9-10 in Fig. \ref{fig4}).

\begin{algorithm}
    \footnotesize
    \SetKwInOut{Input}{Input}
	\SetKwInOut{Output}{Output}
	\Input{string $token$, address $pk$, signature $t$, operation $op$}
    \Output{$out$}
	\textbf{Initialisation:} $rec$ $\leftarrow$ $null$, $out$ $\leftarrow$ $rejected$ \\
	\BlankLine
	$s \leftarrow \mathcal{V}(pk, t)$
	\Comment{$\mathcal{V}$: Signature verifying function} \\
	\If{$s$} {
	    $rec$ $\leftarrow$ $token\_ledger$[$token$] \\
	    \If{(($rec.DO\_ID = pk$) $\lor$ ($rec.SP\_ID = pk$))} {
	        $token\_ledger$[$token$].$expires\_in$ -= Time.now(); \\
	        $token\_ledger$[$token$].$refresh\_count$ += 1; \\
	        $out$ $\leftarrow$ $accepted$
	    } \Else{
	        \If{($rec.TP\_ID = pk$) $\land$ ($rec.permission \subset op$) $\land$ \\
                ($rec.expires\_in > 0$) $\land$ ($rec.status = true$) $\land$ \\ ($rec.refresh\_count < threshold$)}
            {                   
                $token\_ledger$[$token$].$expires\_in$ -= Time.now(); \\
	            $token\_ledger$[$token$].$refresh\_count$ += 1; \\
                $out$ $\leftarrow$ $accepted$\\
            }
        }
	}
	\textbf{Return} $out$
	\caption{$Validation$ invoked by a RS for validating a data query request and updating the $token\_ledger$ accordingly}\label{alg2}
\end{algorithm}
\section{Implementation and Preliminary Results}
We consider a use-case a clinic (\textit{i.e.,} SP) collects clinical trials from patients (\textit{i.e.,} DOs) and stores in a secure and honest RS. This data is then shared with TPs for research. The BC-based clinical data management system is built on top of the Ethereum framework. The source-code is GPL licensed and can be found at our Github\footnote{https://github.com/nguyentb/Personal-data-management}.

\subsection{Resource Server Setup}
We build a RESTful web-service as a RS allowing parties to query clinical trials by calling RESTful APIs. Clinical trials are stored in JSON-like documents using MongoDB\footnote{https://www.mongodb.com/}, a document-oriented database. The trial information includes patientID, name, contact information and data object. Data query includes $\{create, read, update, delete\}$ (\textit{i.e.,} CRUD operations) on clinical trials. An API call to the RS is an HTTP method comprising of $Method$ (\textit{e.g.,} $GET$ and $POST$), $REST.Endpoint$ (\textit{e.g.,} $localhost$:$8080$), $API.Endpoint$ (e.g, $/ClinicalDataManagement$), $Header$ (\textit{e.g.,} $Content.Type$:$application/json$) following by $Params$ including identity $pk$, signature $t$, $access\_token$, and a CRUD operation $op$, along with $payload$ in the body.

The RESTful web-service is required to be an client of the Ethereum private network for the API call validation. An additional Javascript program leveraging the Web3JS library\footnote{https://github.com/ethereum/web3.js} is integrated with the RESTful APIs for interacting with the Ethereum network. The Javascript program proposes transactions invoking the $Validation$ function with necessary parameters provided in an API call.

\subsection{Smart Contracts Implementation}
The SC is implemented in Solidity deployed in an Ethereum test-net using Truffle suite framework with Ganache tool\footnote{https://truffleframework.com}. In the demonstration, Ethereum $address$ is used for party identity. This means patients and research institutes are clients of the Ethereum test-net and are assigned with a corresponding $address$ as ID associated with a private-key. Ledgers in form of value-key pairs are defined as global variables using the built-in $mapping$ type in which "key" is a primitive types and "value" is a user-defined struct. For example, $token\_ledger$ is defined according to the data format in Listing \ref{lst2} as follows:
\begin{small}
\begin{verbatim}
mapping (string => Token) public token_ledger
\end{verbatim}
\end{small}
where "key" in $token\_ledger$ is a string generated by Eq. \ref{eq_2} and "value" is an instance of the user-defined $Token$ struct.

As Solidity only allows primitive types as "key" in $mapping$ and the "key" in the $access\_ledger$ data format defined in Listing \ref{lst1} contains three IDs, we separate the $access\_ledger$ data model into three separated different ledgers namely $ACL\_ledger$, $ACL\_keeper\_ledger$, and $Data\_ledger$. The three ledgers are connected by referring parties' ID (\textit{i.e.,} Ethereum $address$) to each other. In the demonstration, data usage policy is in form of Access Control List defined by a nested mapping between DO $address$, TP $addresses$, and an instance of $ACL$ struct as shown in the below code . The $ACL$ struct consists of a combination of the CRUD operations and an associated $access token$. Based on these clarifications, required functions including $Validation$ are successfully carried out.
\begin{small}
\begin{verbatim}
mapping (address => mapping (address => ACL)) ..
    public ACL_ledger
\end{verbatim}
\end{small}

\subsection{System Performance Evaluation}
\begin{figure}[!htbp]
\centering
\captionsetup{justification=centering}
	\includegraphics[width=0.4\textwidth]{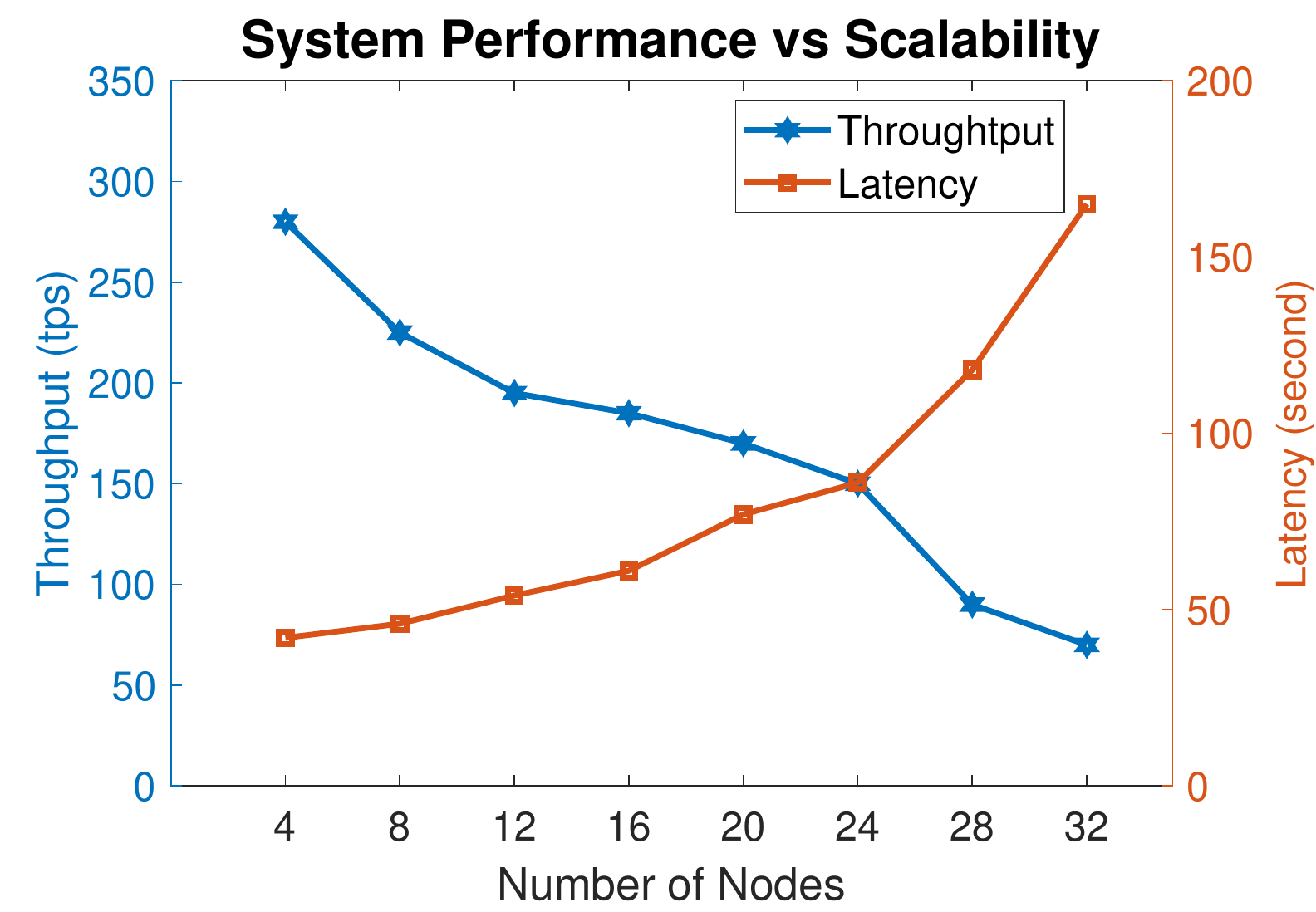}
	\caption{Performance vs Scalability with 8 clients intensively incurring high workload}
	\label{fig5}
\end{figure}

Our aim in the demonstration is to show the feasibility of the proposed model instead of optimising system performance. We carry out a benchmark for our private Ethereum network utilising the BLOCKBENCH tool \cite{ref52}. In the evaluation, a network consisting of varied number of full-nodes from $4$ to $32$ is setup, in which $8$ concurrent clients intensively incur workload to the Ethereum system in every $5$-minute period. Fig. \ref{fig5} interprets the results on the performance and scalability of our system. As shown in the figure, the Ethereum-based system only reaches maximum $375$ transactions per second (tps) with $4$ full-nodes running PoW at about $46$ seconds latency implying that the system only serves limited number of simultaneous requests. Severely, when scaling up, the performance is considerably deteriorated in terms of both throughput and latency. For instance, with the setup of $32$ full-nodes, the Ethereum-based platform can only serves $70$ tps with the delay at around $165$ seconds. The results show that the system is only suitable for small-scale services. As the system performance heavily depends on the underlying BC network instead of an application built on top, getting better performance requires more research on Ethereum protocols such as consensus, sharding and micro-payment, which are out of score of this paper.

\section{Conclusion and Future Work}
In this paper, we point out three crucial flaws in the existing BC-based personal data management systems due to the inadequacy and inefficiency of the design architecture that let a BC network fully handle data access from/to an off-chain storage. We then propose a novel solution to deal with these flaws by re-designing the system model that utilises a novel concept called decentralised "proof of permission" represented in form of \textit{access token}. An \textit{access token} is issued and delivered to a party by a BC network if it meets data usage policy recorded in a ledger. The party then use the \textit{access token} to query data as long as the query request passes the validation from a RS with the BC. The feasibility of the solution is successfully demonstrated by the clinical trials management use-case. We believe this research removes existing ambiguity and plays a catalyst to develop a practical BC-based personal data management.

As a future work, performance evaluation will be conducted comparing different system designs with different BC platforms. Herein, a distributed off-chain storage in which some storage nodes may misbehave should be taken into account. More efforts are needed to resolve the lack of a trusted storage system for a fully decentralised system. An automated fine-grain expressive usage control in which a SC implementing an smart policy generator in a context-aware fashion is a promising research direction. Last but not least, as data queries only relate to storage mindset (CRUD operations only), BC can be exploited for computational capability, meaning BC nodes, as secure multi-party computation agents, carry out data processing and return requested results instead of raw data.

\section*{Acknowledgment}
This research was supported by the HNA Research Centre for Future Data Ecosystems at Imperial College London.

\bibliography{refs}
\bibliographystyle{IEEEtran}

\end{document}